\documentstyle[aps,prl,psfig]{revtex}
\newcommand{\req}[1]{(\ref{#1})}
\newcommand{\be}{\begin{equation}}
\newcommand{\ee}{\end{equation}}
\newcommand{\bea}{\begin{eqnarray}}
\newcommand{\eea}{\end{eqnarray}}

\newcommand{\pr}[1]{\left(#1\right)}

\newcommand{\avg}[1]{\langle{#1}\rangle}
\newcommand{\up}{\uparrow}
\newcommand{\down}{\downarrow}
\newcommand{\sign}{\textrm{sign}\ }
\newcommand{\beas}{\begin{eqnarray*}}
\newcommand{\eeas}{\end{eqnarray*}}
\def\sign{\hbox{sign}\,}
\def\erf{\hbox{erf}\,}

\newcommand{\BE}{\begin{eqnarray}}
\newcommand{\EE}{\end{eqnarray}}
\newcommand{\BEn}{\begin{eqnarray*}}
\newcommand{\EEn}{\end{eqnarray*}}
\newcommand{\barr}{\begin{array}}
\newcommand{\earr}{\end{array}}

\newcommand{\bit}{\begin{itemize}}      
\newcommand{\eit}{\end{itemize}}
\newcommand{\bc}{\begin{center}}
\newcommand{\ec}{\end{center}}
\newcommand{\ben}{\begin{enumerate}}    
\newcommand{\een}{\end{enumerate}}

\newcommand{\ovl}{\overline}
\newcommand{\Tr}{\text{Tr}\ }
\newcommand{\Avg}{\avg}

\newcommand{\om}{\omega}
\newcommand{\Om}{\Omega}

\begin{document}

\twocolumn[\hsize\textwidth\columnwidth\hsize\csname
@twocolumnfalse\endcsname
\title{Modeling Market Mechanism with Minority Game}
\author{Damien Challet$^{(1)}$, Matteo Marsili$^{(2)}$ 
and Yi-Cheng Zhang$^{(1)}$}
\address{ $^{(1)}$ Institut de Physique Th\'eorique, 
Universit\'e de Fribourg, CH-1700\\
$^{(2)}$ Istituto Nazionale per la Fisica della Materia (INFM),
Trieste-SISSA Unit, V. Beirut 2-4, Trieste I-34014,\\
}
\date{\today}
\maketitle
\widetext

\begin{abstract}
Using the Minority Game model we study a broad spectrum of problems of market 
mechanism. We study the role of different
types of agents: producers, speculators as well as noise traders. The central 
issue here is the information flow : producers
feed in the information whereas speculators make it away. How well each agent fares in the common game depends on
the market conditions, as well as their sophistication. Sometimes there is much to gain with little effort, sometimes 
great effort virtually brings no more incremental gain. Market impact is shown to play also an important role, a strategy 
should be judged when it is actually used in play for its quality. Though the Minority Game 
is an extremely simplified market model, it allows to ask, analyze and answer 
many questions which arise in real markets.\\

\end{abstract}

]
\vspace{4ex}

\narrowtext

\section{Introduction}
Recently it became possible to study markets of heterogenous agents, in particular
in the form of the so-called Minority Games (MG)\cite{CZ1,ZEnews}. Since long time practitioners of the market,
as well as some economists have criticized the main-stream economics where a so-called
representative agent plays the central role. Many prominent economists like Herbert Simon \cite{Simon},
Richard Day, and Brian Arthur \cite{Arthur} have been forceful proponents of the "bounded rationality" and
"inductive thinking". However, though many people join in unison in their criticism of the main-stream,
their alternative approaches and models do not command consensus yet. 

The MG is inspired by Arthur's "{\it El Farol}" model \cite{Arthur}, which shows for the first time how the equilibrium
can be reached using inductive thinking. Whereas {\it El Farol} model is about the equilibrium, our MG model
is about fluctuations.
In a sense MG gives us a powerful tool to study detailed pattern of fluctuations, the equilibrium point is trivial by 
design. It is the fluctuations that play the dominant role
in economic activities, like the market mechanism. MG allows us to study in a precise manner how is the
approach
to equilibrium, how the agents try to outsmart each other, for their selfish gain, compete for the
available
marginal information (any deviation from the mid-point represents 
exploitable 
advantage). It is for this residual margin all the agents fight for, resembling the real markets. The importance 
to market mechanism is primordial,
as any practitioner can attest. Neo-classical economics would tell us that the MG, as in a competitive market,
does not offer consistent gain, based on the Efficient Market Hypothesis (EMH). However, if some agents stop
playing (i.e. choosing dynamically among the two sides), they will give off information that the other more deligent
dynamic agents make away. In the sense the equilibrium can be only dynamically maintained, any relaxing would
imply a relative disadvantage. Isn't that the same thing in real markets?

Studying a model of market mechanism opens up many detailed questions, which practitioners have to face
constantly but the main-stream economists do not have any clue to answer them. For instance, in a model
 like MG agents interact with each other through a common market, what information each agent brings in? 
What gain each agent takes out? How sophisticated should an \hbox{agent} be? What is realizable gain objective?
What is the role of noise traders? How about insider trading (an agent processing privileged information
about fellow agents)? What is the market impact of an otherwise clever strategy? The list is obviously
endless. The point we want to make here is that with so little to start with, and with so many questions
relevant to real markets one can hope for a qualitative answer.  

After two years since the MG's birth, during which much work
has revealed its extremely rich structure \cite{web}, an analytical approach leading
to its exact solution has been found \cite{CMZe,MCZ}.
Unfortunately the main progress is still confined
in the physics community. We hope that, with this paper, this will change: The aim is to 
convince people, including economists hopefully, that many concrete questions about market
mechanisms can be asked and answered, in precise and analytical way, using the 
approach of refs. \cite{CMZe,MCZ}. In fact the MG can be used
as a flexible platform and different handles can be added and manipulated almost at wish.
To achieve our goal, the anaytic approach shall be supplemented by numerical simulations
to confirm its validity. More technical parts and heavy calculations shall be dealt with 
in the appendices.

\section{Main results}

Here below we give a list of salient points of our paper:
\begin{enumerate}

\item Diversification of ideas. If an agent has different alternative strategies, the 
latter may have better to be diversified, i.e. not too much correlated. We show the effects of diversification.

\item Markets have two types of agents: producers and speculators. The former do not have alternative
   strategies; the latter are represented by the normal agents of the standard
   MG. Producers provide information into the market, upon which speculators feed. For the first time it is
   possible to demonstrate that producers and speculators need each other, they live in a symbiosis.
   However, benefits to each group are not equal, depending on the parameters. 

\item Agents are not obliged to play, if they do not see a possible gain. We generalize MG to let agents
   to have the option of not playing. In the presence of producers, markets appear to be attractive and
   more speculators are drawn into the fore. 

\item Noise traders.  One may wonder if some traders decide to be pure noise traders, i.e. they use completely
   random strategy, what is the "harm" done to other market participants (producers and speculators), as well
   as to themselves. In the information rich, they appear to increase volatility and in the herding-effect
   phase, they actually make the market perform better.

\item Despite of the fact that agents start equally equipped, there are better and worse agents and the rank
   of the agents has an interesting non Gaussian "bar-code" structure. 

\item Sometimes it pays to increase the capacity of an agent's brain, say add one more unit in $M$. This
   will give enormous advantage to the better equipped agent in the crowded phase (or symmetric phase)
   where information on the range $M$ is exhausted, whereas such a feature becomes a disadvantage
 in the information rich phase.

\item Does it pay to have more strategies as alternatives? In general yes. Here we calculate the relative
   advantage by having more alternatives. We also show that, due to self-market impact, the imagined gain
   differs from the real gain, a fact known too well to market practitioners. Even each agent has 
   many alternatives, they actually use only a small number of them. 

\item Some agents may get illegal information about others. Just like a stock broker who knows his clients'
    orders before execution. Hence he has privileged information and should be barred from trading. One agent 
who spies on fellow agents enjoys trading advantages. We measure how much is this effect, as the
number of fellow
agents whom you spy increases, how much would be your gain.

\end{enumerate}

\section{Formalism and review}
\label{formalism}
Our model of market consists of $N$ agents which, for
simplicity, can take only one of two actions, such as ``buy''
and ``sell'' at each time step $t$. 
We represent this assuming that each agent 
$i=1,\ldots, N$, at time $t$, can either do the action 
$a_i(t)=+1$ or the opposite action $a_i(t)=-1$. Given the actions
of all agents, the gain of agent $i$ is given by:
\be
g_i(t)=-a_i(t) A(t),~~~~\hbox{where}~~~~A(t)=\sum_{j=1}^N a_j(t).
\label{marketint}
\ee
This equation models the basic structure of market
interaction where each agent's payoffs are determined
by the action taken and by a global quantity $A(t)$, which
is usually a price and it is determined by all of them.
For the sake of simplicity, we assume here a linear
dependence of $g_i(t)$ on $A(t)$. Other choices, such
as $g_i(t)=-a_i(t){\rm sign}\, A(t)$ in refs. 
\cite{CZ1,Savit,CZ2},
can be taken without affecting qualitatively 
the results we shall discuss below. This interaction
clearly rewards the minority of agents (those who
took the action $a_i(t)=-{\rm sign}\, A(t)$) who
gain an amount $|A(t)|$ and punishes the majority
by a loss $-|A(t)|$, hence the name minority game\cite{CZ1}.
There are always more losers than winners and 
agents have no way of knowing what the majority
will do before taking their actions.

All agents have access to  public information
which is represented by an integer variable $\mu$ taking 
one of $P$ values. At time $t$ information ``takes the
value'' $\mu(t)$. We shall also call $\mu(t)$ history
since originally this information has been introduced
as encoding the record of the past $M=\log_2 P$ signs
of $A(t)$ with $M$ bits. It has however been shown \cite{cavagna} 
that if $\mu(t)$ is randomly drawn in
$\{1,\ldots,P\}$ one recovers the same results (see
also the discussion in refs. \cite{CMZe,MCZ}). We shall
henceforth consider this second, simpler case.
When having access to some information, agents can behave differently
for different values of $\mu(t)$, eventually because
of their personal beliefs on the impact that
information $\mu(t)$ shall have on the market's 
outcome $A(t)$. Strictly speaking $A(t)$ only
depends on what agents do, so $\mu(t)$ has no direct
impact on the market. However if agents behavior
depends on $\mu(t)$ also $A(t)$ shall depend on it,
and we denote it by $A^{\mu(t)}(t)$. 

How do agents choose actions under information $\mu(t)$?
If agents expect that $\mu(t)$ contains some information
on the market, they will consider {\em forecasting strategies}
which for each value of $\mu$ suggest which action
$a^\mu$ shall be done. There are $2^P$ such strategies, and we
assume, for the time being, that each agent just picks
$S$ such rules randomly (with replacement) from the
set of all $2^P$ strategies. The action of agent $i$ if
she follows her $s^{\rm th}$ strategy and the information
is $\mu$ is denoted by $a_{s,i}^\mu$. Therefore, 
if $s_i(t)$ is the choice made (in a way we 
shall specify below) by 
agent $i$ at time $t$, her action becomes $a_i(t)\to 
a_{s_i(t),i}^{\mu(t)}$ and correspondingly, her gain
[Eq. \req{marketint}] becomes
\be
g_i(t)=-a_{s_i(t),i}^{\mu(t)} A^{\mu(t)}(t),~~~\hbox{where}~~
A^{\mu(t)}(t)=\sum_{j=1}^N a_{s_j(t),j}^{\mu(t)}.
\label{marketint1}
\ee

In this paper, we mainly focus on $S=2$. 
This case contains all the richness of the model and allows
a more transparent presentation. All the results discussed
below can be extended to $S>2$ along the lines of ref. 
\cite{MCZ}. For $S=2$ we can adopt a notation 
where each agent controls a variable $s_i\in\{\down,\up\}$, with the indentification $\up=+1$ and $\down =-1$. 
This is useful to distinguish
strategies $s_i$ from actions $a_i$.
It is convenient to introduce the variables
\be
\om_i^\mu=\frac{a_{\up,i}^\mu+a_{\down,i}^\mu}{2}, ~~~~~~~
\xi_i^\mu=\frac{a_{\up,i}^\mu-a_{\down,i}^\mu}{2}
\ee
With these notations, the action taken by this 
agent in reaction to the history $\mu$ is:
\be
a_{i,s_i}^\mu=\omega_i^\mu+\xi_i^\mu s_i.
\ee
so $\omega_i^\mu$ represents the part
of $i$'s strategies which is fixed, whereas
$\xi_i^\mu$ is the variable part.
We also define $\Omega^\mu=\sum_i\omega_i^\mu$ so that
\be
A^{\mu}(t)=\sum_{i=1}^N a_{i,s_i(t)}^\mu=
\Om^\mu+\sum_{i=1}^N\xi_i^\mu s_i(t).
\ee

Each agent updates the cumulated virtual payoffs of all 
her strategies according to
\be
U_{s,i}(t+1)=U_{s,i}(t)-A^{\mu(t)}(t)\,a_{i,s}^{\mu(t)}
\ee
The quantity $U_{i,s}$ is a ``reliability index'' which 
quantifies the agent $i$'s perception of the success of her
$s^{\rm th}$ strategy. $U_{i,s}(t)$ is the {\em virtual}
cumulated payoff that agent $i$ would have received up
to time $t$ if she had always played strategy $s$ (with
others playing the strategies $s_j(t')$ which they 
actually played at times $t'<t$). {\em Virtual} here 
means that this is not the real cumulated payoff but
rather that {\em perceived} by agent $i$. These differ,
as explained below and in ref. \cite{MCZ}, because 
agents neglect their impact on the market (i.e. the
fact that if they had indeed always played $s$ the
aggregate quantity $A(t)$ would have been different).

Inductive dynamics \cite{Arthur,CZ1} consists in assuming
that agents trust and use their most reliable strategy, 
which are those with the largest virtual score:
\be
s_i(t)={\rm arg}\max_{s\in\{\up,\down\}} U_{i,s}(t).
\label{choice}
\ee
More generally one can consider a probabilistic choice
rule -- the so called {\em Logit} model\cite{logit} --
such that $P(s_i(t)=s)\propto\exp[\Gamma U_{s,i}(t)]$, (see 
 \cite{cavagna2,CMZe,MCZ}). Then Eq. \req{choice} is 
recovered in the limit $\Gamma\to\infty$.
As in ref. \cite{CM}, we find it useful to introduce the
variables $\Delta_i(t)=U_{i,\up}-U_{i,\down}$. Their dynamics 
reads
\be
\Delta_i(t+1)=\Delta_i(t)-A^{\mu_t}(t)\xi_i^{\mu_t}
\label{dit}
\ee
and Eq. \req{choice} becomes
\be
s_i(t)=\sign \Delta_i(t).
\label{choice1}
\ee

\subsection{Notations on averages}

We define the temporal average of a given time dependent 
quantity $R(t)$ as
\be
\avg{R}=\lim_{T\to\infty} \frac{1}{T}\sum_{t=1}^T R(t).
\ee
This quantity can be decomposed into conditional averages
on histories, that is 

\be
\avg{R^\mu}=\lim_{T\to\infty}\frac{P}{T}\sum_{t=1}^T R(t)\delta_{\mu(t),\mu}.
\ee
Note that the factor $P$ and the relation 
$\avg{\delta_{\mu(t),\mu}}=1/P$ imply that $\avg{R^\mu}$ is 
a conditional average. More precisely it is the 
temporal average of the quantity $R(t)$ subject to the 
condition that the actual history $\mu(t)$ 
was\footnote{This implies that the number of iterations must be 
proportional to $P$ in any numerical simulation} $\mu$. 
Finally, averages over the histories $\mu$ of a quantity $R^\mu$ 
are defined as
\be
\ovl{R}\equiv \frac{1}{P}\sum_{\mu=1}^P R^\mu.
\ee

\subsection{Quantities of interest}
\label{quantities}
With these notations, let us now discuss the main quantities
which characterize the stationary state of the system. 
The main free parameter, as first observed in ref. \cite{Savit}, is
\be
\alpha=\frac{P}{N}
\ee
and we shall eventually consider the thermodynamic limit
where $N,P\to\infty$ with $\alpha$ fixed.
The first quantity of interest is 
\be
\sigma^2\equiv\ovl{\avg{A^2}}=
{\ovl{\Om^2}+2\sum_{i=1}^N \ovl{\Omega\xi_i}\avg{s_i}
\avg{s_i}+\sum_{i,j}\ovl{\xi_i\xi_j}\avg{s_i s_j}}.
\ee
This equals the total loss of agents 
\be\label{gainsigma}
-\sum_i \avg{g_i}=\sigma^2,
\ee
so it is a measure of global waste. 
It also quantifies the market's ``volatility'' 
i.e. the fluctuations of the quantity $A(t)$, and is related to 
the average ``distance'' between agents (see appendix \ref{geom_alg}).
Even though
$\avg{A}=0$, by symmetry, it may happen that for a 
particular $\mu$, the aggregate quantity $A(t)$ is nonzero 
on average, i.e. that $\avg{A^\mu}\neq 0$. In order
to quantify this asymmetry, we introduce the quantity
\be
H\equiv\ovl{\avg{A}^2}={\ovl{\Om^2}+2\sum_{i=1}^N 
\ovl{\Omega\xi_i}\avg{s_i}
\avg{s_i}+\sum_{i,j}\ovl{\xi_i\xi_j}\avg{s_i}\avg{s_j}}.
\label{H}
\ee
Note that the only difference with $\sigma^2$ lies in the
diagonal terms ($i=j$) of the last sum. Indeed we 
assume that $\avg{s_i s_j}=\avg{s_i}\avg{s_j}$ for 
$i\ne j$, whereas\footnote{This amounts to say that the 
fluctuations in time of $s_i$ around its average 
$\avg{s_i}$ are uncorrelated with $s_j-\avg{s_j}$. 

This assumption fails when crowd effects occur, i.e. 
in the symmetric phase, and our theory will accordingly
fail to describe these effects.} $\avg{s_i^2}\equiv 
1\neq\avg{s_i}^2$.
Indeed, we can write
\be
\sigma^2=H+\sum_{i=1}^N\ovl{\xi_i^2}(1-\avg{s_i}^2).
\label{relHsig}
\ee
If $H>0$, the game is asymmetric: At least for some $\mu$ 
one has that $\avg{A^\mu}\not =0$. This implies that there
is a {\em best} strategy $a^\mu_{\rm best} =-\sign\avg{A^\mu}$ which 
in principle could give a positive gain $\ovl{|\avg{A}|}-1$
\footnote{Here the $-1$ comes from the fact that if the
strategy is actually played $A^\mu\to A^\mu+a^\mu_{\rm best}$
and ``in principle'' means that $A^\mu$ would also change as a 
result of the fact that other agents would also react to 
the best strategy agent.}. In economic terms we may say
that the system is not {\em arbitrage free}, and that $H$
is a measure of the perceived arbitrage opportunities present
in the market. As a function of $\alpha=P/N$ the system 
displays a {\em phase transition} with symmetry breaking\cite{CM}:
For $\alpha>\alpha_c$ the symmetry between the two signs
of $A(t)$ is broken.

$H$ plays a particular important role because in refs. 
\cite{CMZe,MCZ} it has been shown that the inductive 
dynamics is equivalent to a dynamics which minimizes $H$
in the dynamical variables $m_i=\avg{s_i}$. Therefore the ground
state properties of the Hamiltonian $H$ yields the stationary
state of the system.
$H$ is a spin glass Hamiltonian where $\ovl{\Omega\xi_i}$ 
are the local magnetic fields and $\ovl{\xi_i\xi_j}$ the coupling 
between two agents. These play the same role as {\em quenched}
disorder in spin glasses. This system is of {\em mean field} type
since interactions $\ovl{\xi_i\xi_j}$ are infinite ranged. 
For this reason, the
statistical mechanics approach to disordered systems \cite{MPV,dotsenko}
via the replica method yields exact results for these models
(see appendix \ref{replica}).
The behavior of each agent is completely determined by the
difference of her cumulated payoffs $\Delta_i$. For long times, 
$\Delta_i\simeq v_i t$ where 
\be
v_i=\avg{\Delta(t+1)-\Delta_i(t)}=-2\ovl{\avg{A}\xi_i}. 
\ee
If $v_i\not =0$ agent $i$ shall stick to only one strategy
$s_i=\sign v_i$, whereas if $v_i=0$ she will sometimes use
her $\up$ strategy and sometimes her $\down$ one. This is
quantified by $m_i$, and a global measure of the fluctuations
in the strategic choices of agents is given by
\be
Q=\frac{1}{N}\sum_{i=1}^N m_i^2.
\label{Q}
\ee
This quantity also emerges naturally from the replica approach
where it plays a key role.


\section{Speculators with diversified strategies}

In the standard MG, it is assumed that the agents draw all their strategies randomly, and independently. One can argue that the agents can be less simple-minded so that they first draw a strategy, and then following their needs or what seems the best for them, draw the others strategies. For instance, if $S=2$, an agent can believe that one strategy is enough and sticks to it (or takes two same strategies). Reversely, an agent might believe that it is better to have one strategy and the opposite one. More generally, we suppose that all the agents\footnote{This can be generalized to a $c$ for each agent ; exact results also arise from the replica calculus} draw their second strategy according to

\be
P(a_{\up}^\mu=a_{\down}^\mu)=c~~~\forall \mu.
\ee

\begin{figure}
\centerline{\psfig{file=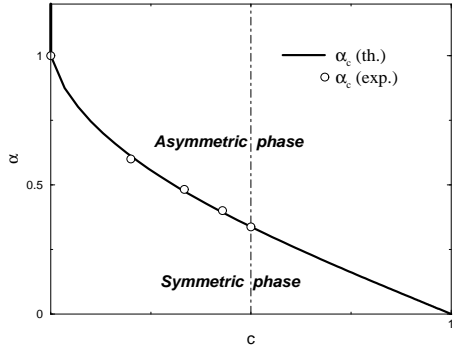,width=6cm}}
\caption{Phase diagram of the Minority Game with diversified strategies. The phase transition in the standard MG corresponds to the dash-dotted vertical line $c=1/2$. The circle are numerical data}
\label{phasediagrC}
\end{figure}

The parameter $c$ counts the average fraction of histories for which the agents' choices are biased, that is, the average correlation between their two strategies. The standard MG corresponds to the independent case $c=1/2$, while having only one strategy is obtained with $c=1$. The other very special case is $c=0$ : all agents have two opposite strategies, thus there is no asymmetry in the outcome. As a result, the game is always in the symmetric phase : as $\alpha$ is varied, no phase transition occurs. Increasing $c$ has two effects : on one hand it increases the bias of the outcome $\Omega^\mu\sim \sqrt{cN}$, on the other hand it reduces the ability of the agents of being adaptative, since they learn something about the game only when $\xi_i^\mu\ne0$ (see Eq \req{dit}), which happens in average for $(1-c)P$ histories. The fact that the biases depend on $c$ too implies that the second order phase transition also occurs when this parameter is varied. With the replica formalism (see appendix \ref{replica}), one gets the phase diagram of the MG with parameter $c$ (see figure \ref{phasediagrC}). In the standard MG, one varies $\alpha$ (dot-dashed vertical line). If one fixes $\alpha$ and changes $c$, the symmetry is also broken (any horizontal line). Note that if $c=0$ and $\alpha>1$, an infinitesimal $c$ breaks the symmetry of the game.

\section{Speculators and producers}

Real markets are not {\it zero sum} games \cite{ZMEM}. The fact that most participants are interested in playing is beyond doubt.
In real markets the participants can be grossly divided into two groups: Speculators and Producers \cite{ZMEM}. Producers can be characterized by those using the market for purposes other than speculation. They need market for 
hedging,  financing, or any ordinary business. They thus pay less or no attention to "timing the market". Speculators, on the
other hand, join the market with the aim of exploiting the marginal profit pockets. The two groups were shown to live
in symbiosis \cite{ZMEM}: the former inject information into the market prices, and the latter make a living carefully
exploiting this information. One may wonder why do producers 
let themselves be taken advantage of.
Our answer is that they have other, probably more profitable business in mind. To conduct their business, they need 
the market, and their expertises and talents in other areas give them still better games to play. Speculators, being
less capable in other areas, or by choice, make do exploiting the "meager margin" left in the competitive market.

In our MG, these general questions can be studied in detail. Producers will be limited in choice, their activities
outside the game are not represented. 
We define a {\bf speculator} as an normal agent, and a {\bf producer} as an agent limited to one strategy. Thus the latter
have a fixed pattern in their market behavior and put a measurable amount of information into the market, which is exploited by the speculators. We take a population of $N$ speculators and always define $\alpha=P/N$. We add $\rho N$ heterogeneous producers, so that $\rho$ is the fraction of producers per speculator. The outcome is then

\be
A^\mu=A_{\text{spec}}^{\mu}+A_{\text{prod}}^{\mu}.
\ee

\begin{figure}
\centerline{\psfig{file=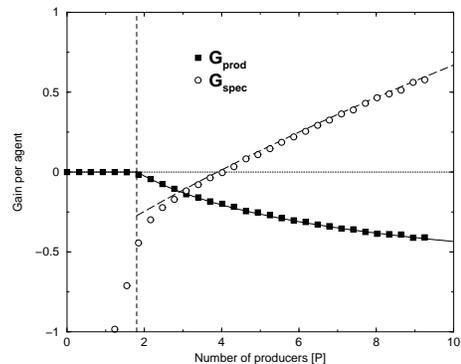,width=6cm}}
\caption{Gain of producers and speculators versus the number of producers (in $P$ unit); the number of speculators is fixed at $N=641$ ($c=0$, $M=8$, $S=2$, $\alpha=0.4$, average over 200 realizations). The lines are theoretical predictions.}
\label{gainprodspecNp}
\end{figure}

The bias induced by the producers adds to the one caused by the speculators, so that the total bias is of order $\sqrt{(c+\rho/2)N}$. Therefore the phase transition can be obtained at fixed $P$ by varying either $N$, $c$, or the number of producers. Let us begin with the last possibility. We fix $c=0$, $P=2^8$, $N=641$ and plot the gains of the speculators and producers as a function of the number of producers (see figure \ref{gainprodspecNp}). In the symmetric phase, the speculators wash out all the available information, thus, by symmetry, the gain of the producers (squares) is zero. As the number of producers increases, the gain of the speculators (circles) stays negative but grows monotonically, while the gain of the producers remains zero as long as the symmetry of the outcome is not broken. When the number of producers reaches a critical value, the speculators are no more able to remove all the available information, therefore the (second order) phase transition occurs (dashed line). Beyond this point, the producers lose more and more, while some (frozen) speculators gain more than zero in average (see \ref{gain}). At one point, the gains of speculators and producers are the same. Finally, there are enough producers to make the gain of the speculators positive on average.

\begin{figure}
\centerline{\psfig{file=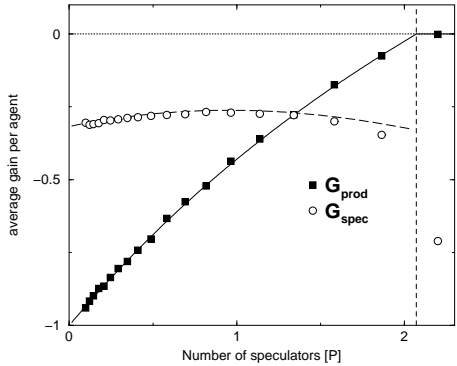,width=6cm}}
\caption{Gain of producers and speculators versus the number of speculators (in $P$ unit); the number of producers 
is fixed at $64$ ($c=0$, $M=8$, $S=2$, average over 200 realizations). The lines lines are theoretical predictions.}
\label{gainprodspecNsbis}
\end{figure}

\begin{figure}
\centerline{\psfig{file=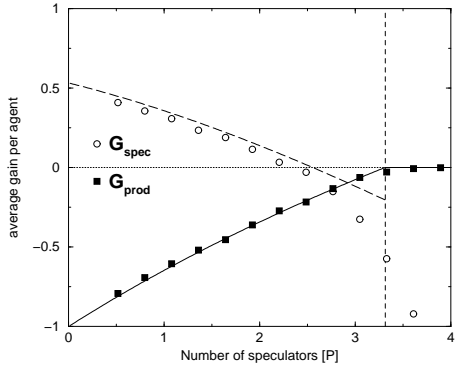,width=6cm}}
\caption{Gain of producers and speculators versus the number of speculators (in $P$ unit); the number of producers 
is fixed at $256$ ($c=0$, $M=6$, $S=2$, average over 200 realizations). The lines are theoretical predictions.}
\label{gainprodspecNs}
\end{figure}

As illustrated by figures  \ref{gainprodspecNsbis} and \ref{gainprodspecNs}, if the number of speculators changes
the behavior is qualitatively the inverse of the one of figure \ref{gainprodspecNp}:
The gain of producers increases as the number of producers grows; similarly, the gain of the speculators 
decreases when $N$ increases for sufficiently large $N$. If there are not enough producers, the game is 
always negative sum for the speculators, and their gain has a maximum (see figure \ref{gainprodspecNsbis}). 

We now expose exact analytical results concerning the gain of the two types of agents. 
They rely on the generalization of the approach of refs. \cite{CMZe,MCZ}: the 
calculus is carried out in detail in appendix  \ref{replica}.
Let us introduce $G_{\text{spec}}$, the total gain of the speculators and $G_{\text{prod}}$, 
the one of the producers. From Eq \req{gainsigma}

\be
G_{\text{spec}}+G_{\text{prod}}=-\sigma^2.
\ee

The results depend on the ratio  $\rho$ between the number of 
producers, on the number of speculators and on $c$, the parameter introduced in the previous section. We obtain
\be
\frac{\sigma^2}{N}=\frac{c+\rho+(1-c)Q}{(1+\chi)^2}+(1-c)(1-Q)
\ee  
where $\chi$ is the magnetic susceptibility of the system, 
and $Q$ is defined in section \ref{quantities}. 
These two quantities depend on $\alpha$ and on $(1+\rho)/(1-c)$ (see appendix  \ref{replica}). 
The average gain per producer is 
\be
\frac{G_{\text{prod}}}{\rho N}=-\frac{1}{1+\chi}
\ee
and the average gain per speculator is
\be\label{gainspec}
\frac{G_{\text{spec}}}{N}=
-\frac{c+\rho+(1-c)Q}{(1+\chi)^2}-(1-c)(1-Q)+\frac{\rho}{1+\chi}.
\ee

Figures \ref{gainprodspecNp}, \ref{gainprodspecNsbis} and \ref{gainprodspecNs} completely agree with analytical results; note that the small deviations are finite size effects.
The fact that the gains of producers and speculators only depend on the ration $\rho$ and not on how many producers and speculators there are in the game explains why figures  \ref{gainprodspecNsbis} and \ref{gainprodspecNs} look very much like the inverse of figure \ref{gainprodspecNp}.

\begin{figure}
\centerline{\psfig{file=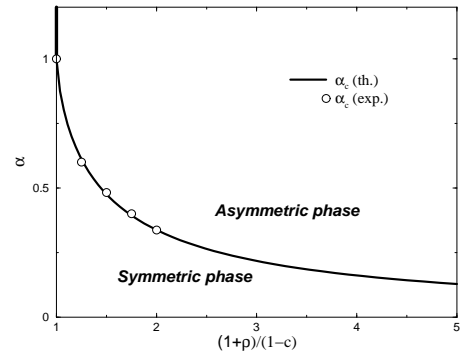,width=6cm}}
\caption{Phase diagram $\alpha_c[(1+\rho)/(1-c)]$}
\label{phasediagr}
\end{figure}

\begin{figure}
\centerline{\psfig{file=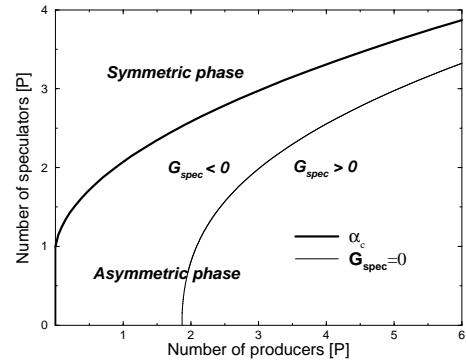,width=6cm}}
\caption{Phase diagram, and zero sum gain for speculators with $c=0$}
\label{phasediagrNp}
\end{figure}

As it emerges for the replica calculus, the critical point $\alpha_c$ only depends\footnote{This explain why evolutionary schemes that preserve the distribution of the quenched disorder have the same $\alpha_c$ \cite{SavitEv}, while others that involve Darwinism, shift $\alpha_c$ \cite{CZ1,CZ2}.} on $(1+\rho)/(1-c)$ (see figure \ref{phasediagr}), that is, on the distribution of the quenched disorder. Numerical data (circles) completely agree with our results. The vertical line corresponds to the standard MG ($\rho=0$ and $c=1/2$). A more intuitive version of this phase diagram is shown in figure \ref{phasediagrNp} for $c=0$.

The game becomes favorable, on average, for the speculators when their average gain is greater than zero. Using Eq \req{gainspec}, one can plot the curve of zero sum gain for the speculators (see figure \ref{phasediagrNp}). One can see that the number of producers must be greater than $1.868\ldots P$ (this value depends on $c$) in order to make the game positive sum for the speculators; this is consistent with numerical simulations (figures \ref{gainprodspecNsbis} and \ref{gainprodspecNs}).

\begin{figure}
\centerline{\psfig{file=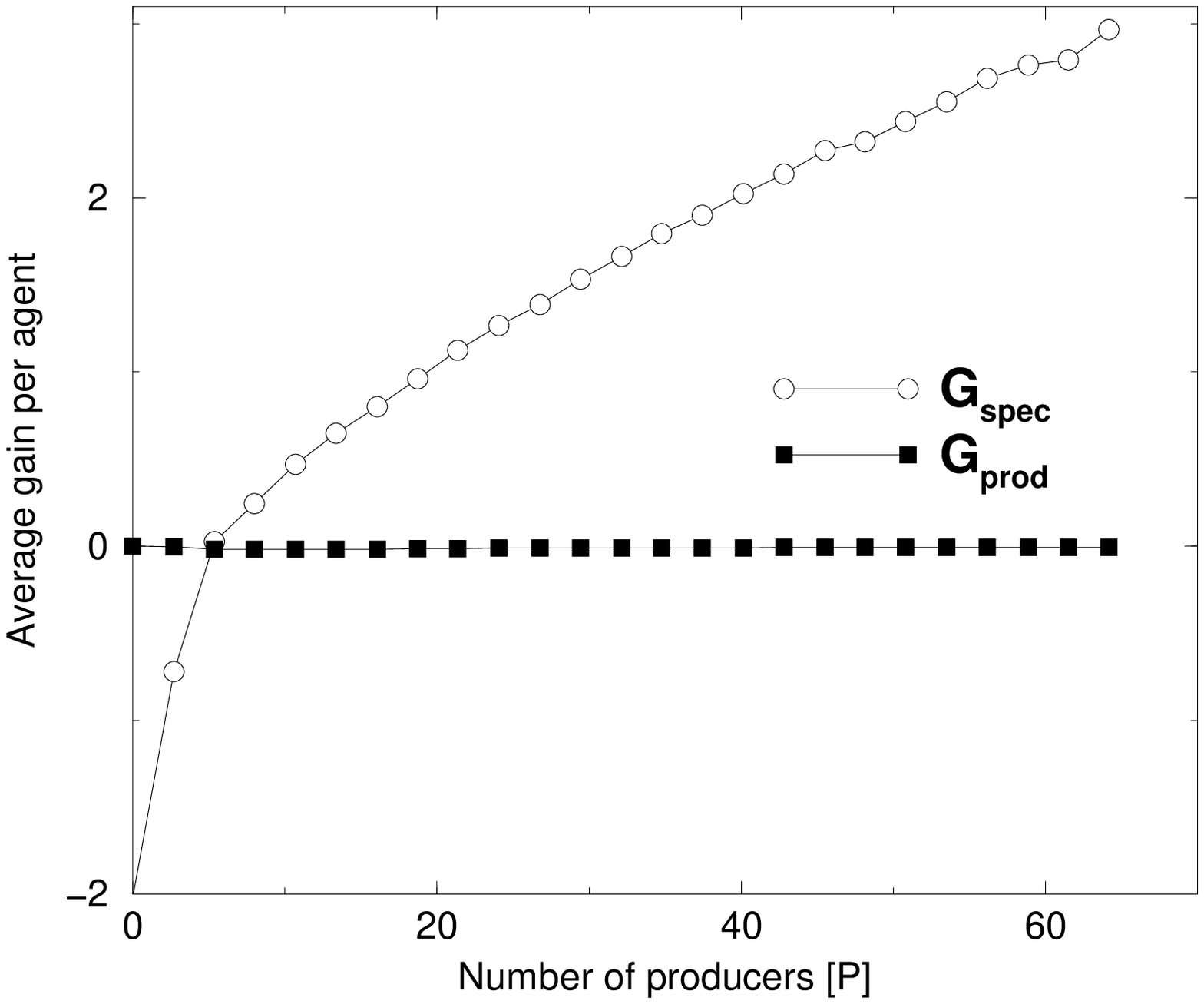,width=6cm}}
\caption{Average gain per agent versus the number 
of producers (in $P$ units) in the grand canonical MG ($N=107$, $M=5$, $\alpha=0.3$, $S=2$, $c=1/2$, average over 500 realizations)}
\label{grandcangain}
\end{figure}

\begin{figure}
\centerline{\psfig{file=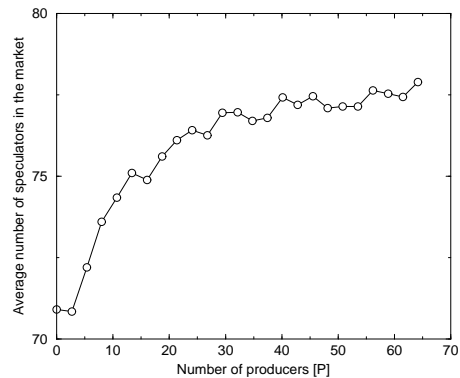,width=6cm}}
\caption{Average number of speculators versus the number 
of producers (in $P$ units) in the grand canonical MG ($N=107$, $M=5$, $\alpha=0.3$, $S=2$, $c=1/2$, average over 500 realizations, )}
\label{grandcan}
\end{figure}

The main message of these results is that producers always
benefit from the presence of speculators, and reversely : both types of agents
live in symbiosis.
 Indeed, the producers introduce 
systematic biases into the market, and without speculators, their
losses would be proportional to theses biases. The speculators precisely try to remove
this kind of bias, reducing also systematic fluctuations in the market, thus reducing the losses
of the producers and their own losses.  Moreover,
the efforts of speculators yield a positive gain only if 
the number of producers is sufficiently large. In this respect
the symmetric phase, where producers do not loose and speculators 
loose a lot, is unrealistic: real speculators would rather 
withdraw from a market which is in this phase, thus increasing 
$\alpha$, and recovering the asymmetric phase. This suggests 
that a grand-canonical MG is much more realistic
\footnote{See also \cite{MZe,JohnsonGR}}. Here we 
briefly present an over-simplified ``grand-canonical'' MG. An agent
 enters into the market only when she has a strategy with virtual 
points greater than zero. As a result, the game is always in the asymmetric phase, but almost at the transition point :
 the average losses of the producers are always extremely small (see figure \ref{grandcangain}). 
When the number of producers increases, the {\em a priori}
asymmetry of the outcome increases, and more and more agents actually 
play the game (see figure \ref{grandcan}), thus in this situation, the producers
give incentives to play to the speculators. Accordingly, 
the average gains of the speculators is much higher in this 
grand-canonical MG than in the corresponding canonical MG.

\section{Speculators, producers and noise traders}

The debate about what the noise 
traders do to a competitive market is not closed \cite{Noise}.
In the economics literature a noise trader is not very precisely defined. 
Sometimes they are synonym with
speculators. We define noise traders in the following way: 
they choose their actions without any basis. Compared with
speculators, who analyze carefully the market 
information, noise traders take action in a purely random way
(see appendix \ref{replica}). Noise traders may be speculators 
who base their action on astrology, on ``fengshui'', or on some 
``random number generators''. 
Our present model allows us to evaluate the
influence of noise traders on the market.
They increase the market volatility $\sigma^2$,
as shown in Fig. \ref{noise} and in appendix  \ref{replica}. 
Therefore, in principle, they do harm 
to themselves as well to other participants. Actually in the
linear--payoff version that we consider, the average gain of 
speculators and producers is not much affected by noise traders, since
$\avg{A_{\rm noise}}=0$. However, it is easy to see that 
in the original version, where $g_i(t)=-a_i(t)\sign A(t)$, 
payoffs are reduced by the presence of noise traders (see 
appendix \ref{sign}). 

Our numerical results of Fig. \ref{noise} also shows that deep 
in the symmetric phase, noise traders reduces the volatility 
per agent $\sigma^2/(N+N_{\rm noise})$, when this becomes bigger 
than one. This is easy to understand assuming that the only effect of
noise traders is to increase $\sigma^2$ by a constant equal to
$N_{\rm noise}\equiv \eta N$. Let $\sigma^2_0/N$ be the 
volatility per agent, without noise traders ($\eta=0$) and
$\sigma^2_\eta$ that with noise traders. The
variation in the volatility per agent in the presence of
noise traders is:
\be
\frac{\sigma_\eta^2}{N(1+\eta)}-\frac{\sigma_0^2}{N}
\simeq \frac{\sigma^2_0+\eta N}{N(1+\eta)}-\frac{\sigma_0^2}{N}
=\frac{1-\sigma^2_0/N}{1+1/\eta}.
\ee
As illustrated by figure \ref{noise}, numerical simulations globally
 confirm these conclusions, but also show that the effects of the noise traders
are more pronounced than the theory predicts.

\begin{figure}
\centerline{\psfig{file=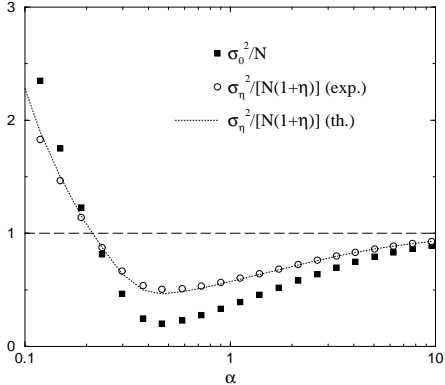,width=6cm}}
\caption{Normalized variance of the outcome with (opaque circles and without (black squares) noise traders; the dotted line is the naive theoretical prediction. Inset : difference of variances with and without noise traders ($N=101$ speculators, $50$ noise traders, average over 1000 realizations).  }
\label{noise}
\end{figure}

\section{Market impact}
\label{market_impact}

In order to quantify the impact of an agent on the market 
let us first consider the case of an {\em external} agent with $S$ strategies:
This agent does not take part in the game but just observes
it from the outside. From this position, each of her strategies
gives an average\footnote{The average is meant over a long time
here} {\em virtual} gain
\be
u_s=-\ovl{a_s\avg{A}}, ~~~~~s=1,\ldots,S.
\ee
Given that the strategies $a^\mu_s$ are drawn randomly, 
$u_s$ are independent random variables. Since $u_s$ is the sum
of $P\gg 1$ independent variables $a_s^\mu\avg{A^\mu}/P$, their
distribution is Gaussian with zero mean and variance
\[
{\rm Var}(u_s)=\frac{1}{P^2}\sum_{\mu=1}^P{\rm Var}(a^\mu_s)
\avg{A^\mu}^2 =\frac{H}{P}. 
\]
Clearly, one of these strategies, that with $u_{s^*}=\max_s u_s$,
is superior to all others\footnote{The distribution of 
$u_{s^*}$ can be easily computed using extreme statistics.
For $S\gg 1$ typically $u_{s^*}\simeq\sqrt{2H\log (S)/P}$.}. 
It would be most reasonable for this agent to just stick to
this strategy.

However, the same agent {\em inside} the game will typically
use not only strategy $s^*$. This is because every strategy, when
used, delivers a {\em real} gain which is reduced with respect 
to the virtual one by the ``market impact''. Imagine the 
``experiment'' of injecting the new agent in a MG.
Then $\avg{A^\mu}\to \avg{A^\mu}+a^\mu_s$, where, in a first
approximation, we neglect the reaction of other agents to the
new-comer. Then the real gain of the newcomer is:
\be
g_s\cong -\ovl{a_s\avg{A}}-\avg{a_s\,a_s}=u_s-1.
\ee

The agent will then update the scores $U_{s}(t)$ with the
real gain $g_s$ for the strategy she uses and with the
virtual one $u_{s'}=g_{s'}+1-\ovl{a_s a_{s'}}$, for the strategies she does not
use (in the following, we neglect the term $\ovl{a_s a_{s'}}$). Therefore inductive agents over-estimate the 
performance of the strategies they do not play.
Then if strategy $s$ is played with a frequency 
$p_s$, the virtual score increases {\em on average} by
\bea
\delta U_s&=&U_s(t+1)-U_s(t)=p_s g_s+(1-p_s)(g_s+1)\nonumber\\
&=&g_s-p_s+1
\eea
at each time step (on average).
If the agent ends up playing only $n$ out of her $S$ strategies 
with some frequency $p_s>0$, it must be that the virtual score
increases $\delta U_s$ are all equal for these strategies and the 
virtual scores of strategies not played is lower. More precisely let
$s=1,\ldots,n$ label the strategies which are played and 
$r=n+1,\ldots,S$ those which are not played. It must be that
\bea
\delta U_s&=&g_s-p_s+1=v~~~~~~~~s=1,\ldots,n\\
\delta U_r&=&g_r+1<v~~~~~~~~~~r=n+1,\ldots,S.
\label{deltaU}
\eea
These equations yield the number $n$ of strategy that this agent
will use. Normalization of $p_s$ in the first equation yields
the average virtual gain $v$ of the agent, which is 
\be
v=\frac{1}{n}\sum_{s=1}^n g_s-\frac{1}{n}+1.
\ee
Using $p_s=g_s+1-v$, we can compute the real gain of the inductive 
agent $g=\sum_{s=1}^n p_s g_s$. 

Summarizing, we find that inductive agents mix their best strategy
with less performing ones. This is a consequence of the fact that they
neglect their impact on the market. 

\begin{figure}
\centerline{\psfig{file=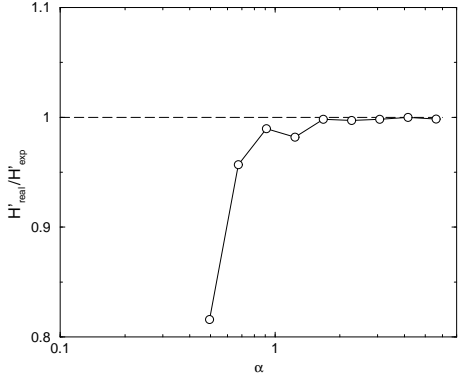,width=6cm}}
\caption{Ratio of real $H'$ over approximated $H'\simeq H-2\ovl{|\avg{A}|}+1$ 
versus $\alpha$ ($N=101$, average over 100 realizations)}
\label{HH}
\end{figure}

So far we did not take into account the reaction of other agents to the new-comer. 
In order to quantify this effect, let us consider a MG in the asymmetric 
phase, and let us add a new agent with the {\em best} strategy $a^\mu=-\sign
\avg{A^\mu}$. This gives us an idea of this effect in the extreme case and we 
expect that for a randomly drawn strategy the effect will be smaller. Neglecting the
reaction of other agents, we find that the available information with the new-comer should be
$H'\simeq H-2\ovl{|\avg{A}|}+1$. Figure \ref{HH} shows that the reaction of all agents is 
indeed negligible, excepted near the critical point, where $H$ is of the order of 1. 

\section{Gain}
\label{gain}

In this section we show how the behavior and the gain of each agent (speculator as well as producer) depends 
on her microscopic constitution and on the asymmetry of the outcome $A(t)$ in the asymmetric phase. Let us denote 
the gain of agent $i$ by $g_i$ ; by definition, 

\be\label{gainbrut}
g_i=-\ovl{\avg{Aa_i}}.
\ee

In the asymmetric phase, since the stationary state is mean field, $\avg{s_i s_j}=m_i m_j$. Consequently, by expanding Eq. \req{gainbrut} one obtains

\bea\label{gaingen}
g_i&=&-\ovl{\avg{A}\omega_i}-\ovl{\avg{A s_i}\xi_i}\nonumber\\
	&=&-\ovl{\avg{A}\omega_i}-\ovl{\avg{A}\xi_i} m_i-\ovl{\xi_i^2}(1-m_i^2).
\eea

Remember that the stationary behavior of agent $i$ is described by 
$v_i=-2\ovl{\avg{A}\xi_i}$ (see section \ref{formalism}). If an agent is non frozen, 
$v_i=0$, while $m_i=-\sign v_i$ otherwise, hence the gain of a generic 
agent $i$ is

\be\label{gainv_i}
g_i=-\ovl{\avg{A}\omega_i}+|\ovl{\avg{A}\xi_i}|-\ovl{\xi_i^2}(1-m_i^2).
\ee
Note that the second term of above equation vanishes for a non frozen agent $j$
and therefore

\be\label{gainactive}
g_j=-\ovl{\avg{A}\omega_j}-\ovl{\xi_j^2}(1-m_j^2)~~~~~\hbox{non frozen}.
\ee
On the other hand, the third term of Eq. \req{gainv_i} vanishes if 
 agent $k$ is frozen:

\be\label{gainfrozen}
g_k=-\ovl{\avg{A}\omega_k}+|\ovl{\avg{A}\xi_k}|~~~~~\hbox{frozen}.
\ee

\begin{figure}
\centerline{\psfig{file=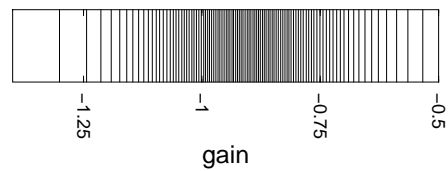,width=6cm,angle=-90}}
\caption{Bar-code structure of the r-th best agent's gain ($\alpha=10$, $M=10$, $S=2$, average over 300 realizations)}
\label{gap}
\end{figure}

In Eqs  \req{gainactive} and \req{gainfrozen}, the gain of each agent is expressed as 
her internal constitution, allowing us to interpret  
what does the gain of a general agent depends on. 
In both equations, the first term $-\ovl{\avg{A}\omega_i}$, which 
represent how much the agents loose due to their bias, is on average 
negative, due to the impact this bias has on the market. 
The second term in Eq \req{gainactive} is always negative, and 
represents the losses due to the switching between strategies,
which, as shown above, arises from the neglect of market impact.
Since the probability distribution function of $m_i$ is not 
Gaussian\cite{CMZe}, 
this term gives rise to an non Gaussian distribution of $g_j$ 
for non frozen agents.
The average gain of the r-th best agent is represented 
in figure \ref{gap}.

\begin{figure}
\centerline{\psfig{file=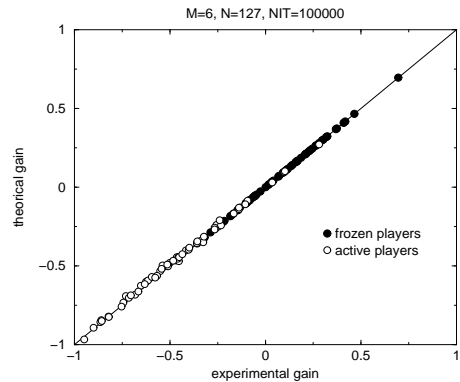,width=6cm}}
\caption{Theoretical gain versus experimental gain showing that the frozen agents gain more than the active ones ($\alpha=0.5$, $M=6$)}
\label{gainthexp}
\end{figure}

By contrast, the term $\ovl{\xi_k^2}(1-m_k^2)$ disappears for a frozen 
agent because $m_k^2=1$. It is replaced by $|\ovl{\avg{A}\xi_k}|$ which is
always positive and which measures how well agent $k$ exploits the available 
information. Therefore, in average, the frozen agents gain more than the non 
frozen ones.  This is clearly illustrated in figure \ref{gainthexp} which also 
shows that Eqs. \req{gainactive} and \req{gainfrozen} are exact. Finally, 
 a producer is of course frozen, and her gain is always zero, since she has
 $|\ovl{\avg{A}\xi_k}|=0$. 

\section{Privileged agent or insider-trading}

In this section we consider a MG where a particular agent
has different characteristics. In particular we address the
question of what additional resources would be advantageous 
for this agent and in which circumstances. 
In the first subsection, we consider an agent with $S'$ strategies
(with $S'>S$, where $S$ is the number of strategies assigned 
to other agents). The last two subsections are devoted to the 
study of effects of asymmetric information, in which an agent 
has access to privileged information which the other cannot 
access. This can be achieved in several ways. First, we consider 
the case of a pure population with memory $M$ and one agent with 
a longer memory $M'$. Then we consider the case of an agent 
who knows, in advance, how a subset of agents plays.

\subsection{An agent with $S'$ strategies}

In the symmetric phase, no matter how many strategies an
agent has, there is no possibility of gaining. Therefore
we focus in this section on the asymmetric phase.

As shown in Sect. \ref{market_impact}, inductive agents over-estimate 
the performance of the strategies they do not play

Let us consider now the case where an agent with $S'$ strategies
enters into a MG. As shown in Sect. \ref{market_impact}, to a
good approximation, the value of $H/P$ is the only relevant information we need to 
retain of the stationary state of the MG without the special agent. 
This quantity encodes all other informations such
as the number of producers, the number of strategies
played by the agents in the MG and the value of $\alpha$.

\begin{figure}
\centerline{\psfig{file=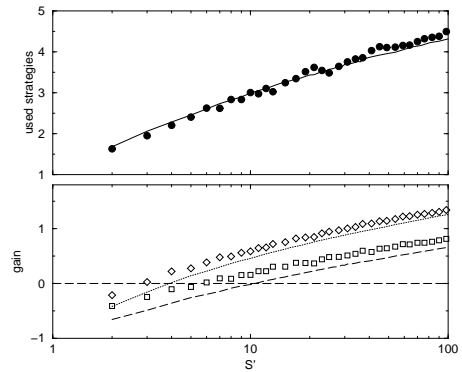,width=6cm}}
\caption{Upper graph : average number of played strategies (circles) versus $S'$. 
Below :  average virtual (diamonds) and actual (squares) gains versus $S'$ for 
$H/P=0.5$, from top to below (averages over 500 realizations). The lines are theoretical predictions}
\label{gain_figfalpha.5}
\end{figure}

\begin{figure}
\centerline{\psfig{file=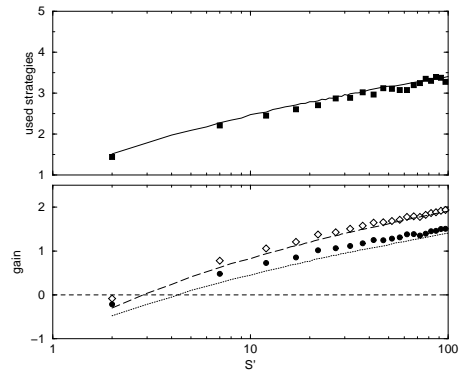,width=6cm}}
\caption{Upper graph : average number of played strategies (circles) 
versus $S'$. Below :  average virtual (diamonds) and actual (squares) 
gains versus $S'$ for $H/P=1$, from top to below (averages over 500 realizations). The lines are 
theoretical predictions.}
\label{gain_figfalpha1}
\end{figure}

We carried out numerical simulations, and compared it
to the analytical results derived in Sect. \ref{market_impact}.
These are shown in the figures 
\ref{gain_figfalpha.5}, for $H/P=0.5$, and 
\ref{gain_figfalpha1}, for $H/P=1$. 
The virtual gain $v$ is always
larger than the actual gain $g$. Even though $g$ is less than
the gain agents would get playing only their best strategy
$E[g_{s^*}]$ (maximal gain), it is not much smaller
and has the same leading behavior $g\propto\sqrt{\ln S}$.

Numerical simulations agree well with analytical results, 
apart from finite size effects which become more pronounced
if $H/P$ is small\footnote{This is mostly due the term which we have neglected in the section \ref{market_impact} : it is typically of the order of $P/H$}.

Figures \ref{gain_figfalpha.5} and \ref{gain_figfalpha1} refer
to values of $H/P$ which are realistic of MG with producers. A
moderately large $S'$ suffices to obtain a positive gain $g>0$.
With $S=2$ and without producers $H/P\sim 0.1$ at most. For these
values the analytic approach suggests that, even playing only 
her best strategy an agent 
would need $S'>750$ strategies to have a positive gain, whereas
inductive agents would need more than $S'\simeq 2400$ 
strategies to obtain a positive gain. The same agent would
find that her virtual gain becomes positive with only $S'>8$ 
strategies. These results for $H/P=0.1$ suffer from strong
finite size effects (which indeed are of the order of $P/H$).
One would need system sizes $N$ which are well beyond what our
computational resources allow to confirm these conclusions.

It is also interesting to observe that the number of strategies 
actually used by the inductive agent increases with $S$ (sub-linearly)
and it decreases as $H/P$ increases (see figures \ref{gain_figfalpha.5} 
and \ref {gain_figfalpha1}). 
That means that if there is
more exploitable information in the system, agent's behavior becomes
more peaked on the best strategy.


\subsection{$M'>M$}

\begin{figure}
\centerline{\psfig{file=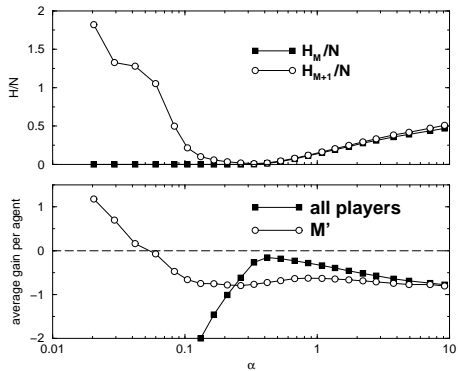,width=6cm}}
\caption{Upper graph : normalized available information for $M$ and $M+1$. Lower graph : Gain of an agent with $M+1$ within a pure population with $M=3$ ($S=2$, average over 3000 realizations) }
\label{bigbrain}
\end{figure}

Let us consider the case of a pure population with memory $M$ 
and one agent with a longer memory\footnote{In this kind of numerical
simulations, one has to keep the dynamics of histories} $M'$. Figure \ref{bigbrain} plots
the gain of such an agent with $M'=M+1$ as a function of $\alpha$.
The average gain of all agents is also shown for comparison.
In the asymmetric phase the special agent receives a lower payoff,
which can be understood by observing that she has a 
number of histories $P'=2^{M'}=2P$ bigger than that of the pure 
population. Thus her effective $\alpha'=2\alpha$ is larger, which is 
detrimental in the asymmetric phase.

The gain of the special agent is the same as
that of normal agents at the point where there is neither persistence, 
nor anti-persistence
($\alpha\simeq 0.25$ for $M=3$, and $\alpha_c$ in the thermodynamic limit). 

By contrast, in the symmetric phase, the game is symmetric for 
normal agents but their anti-persistent behavior produces arbitrages 
who can be exploited by agents having a bigger memory. 
Indeed, as $\alpha$ decreases, the available information 
$H_{M'}$ for the privileged agent grows\footnote{$H_{M'}$ is defined 
as $H=\ovl{\avg{A}^2}$, but with an average over $\mu'=1,\ldots,2P$.}.  
As a result the gain of the privileged agent becomes larger
than that of other agents and as $\alpha$ becomes small enough,
it becomes positive. 

\begin{figure}
\centerline{\psfig{file=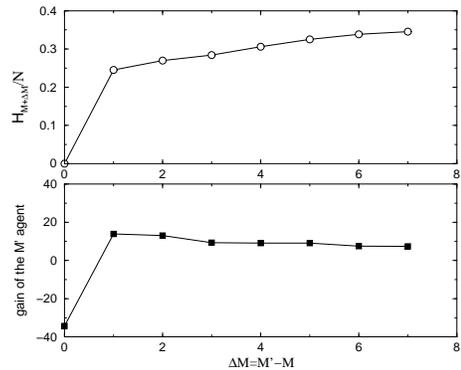,width=6cm}}
\caption{Gain of an agent with $M'=M+\Delta M$ within a pure population with $M=3$ ($\alpha=0.1$, average over 1000 realizations}
\label{bigbrainM}
\end{figure}

Can the anti-persistence be exploited even more if one increases $M'$ ? 
The figure \ref{bigbrainM} answers clearly no. This is not surprising 
since again the effective $\alpha$ is bigger and bigger as $M'$ is 
increased. At the same time, the available information increases, 
but too slowly. 
 
\subsection{Espionage}

Some agents may have access to some information about other
agents. This is the case of a stock broker who knows his clients' 
orders before execution, 
hence he has privileged information and should be barred from 
trading. When there is no available information, as in the symmetric 
phase, an agent who has access to asymmetric information can expect 
at least to lose much less than the others agents, or even to have a 
positive gain. Also, since having access to a little information is 
greatly preferable to no information at all, only a very limited amount 
of information is needed to get a considerable advantage.
Suppose that agent $b$ knows the sign $s_{\cal B}$ of the aggregate 
actions of a subset ${\cal B}$ of other agents. Let $B=|\cal B|$ 
be the number of agents in ${\cal B}$. Then $s_{\cal B}(t)=\sign 
\sum_{i\in{\cal B}} a_{i}(t)$.
She can exploit this supplementary information by having two virtual 
values $U_{b,s}^+(t)$ and $U_{b,s}^-(t)$ for each of her strategies. 
In other words, if agent $b$ knows 
that $s_{\cal B}(t)=+1$ before having to choose, she takes her 
decision according to the scores $U_{b,s}^+(t)$, that is, 

\be
s_b(t)=\arg \max_{s=1,\cdots,S}U_{b,s}^+(t) ;
\ee
she updates the scores of her strategies according to
\be
U_{b,s}^+(t+1)=U_{b,s}^+(t)-a_{b,s}^{\mu(t)}A^{\mu(t)}
\ee
and analogously if $s_{\cal B}(t)=-1$. 

What is the kind of the supplementary information this agent has 
access to ?

\begin{figure}
\centerline{\psfig{file=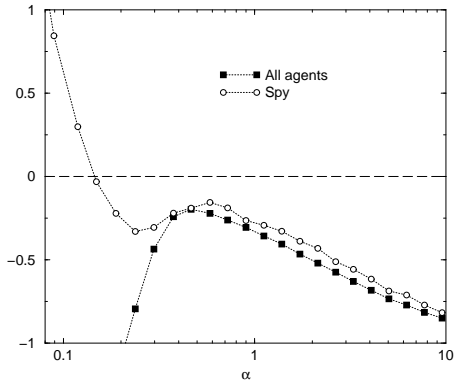,width=6cm}}
\caption{Gain of a spy and average gain of all agents versus $\alpha$ ($N$=101, $N_B=3$, $100 P$ iterations, average over 100 realizations)}
\label{brokeralpha}
\end{figure}

\begin{figure}
\centerline{\psfig{file=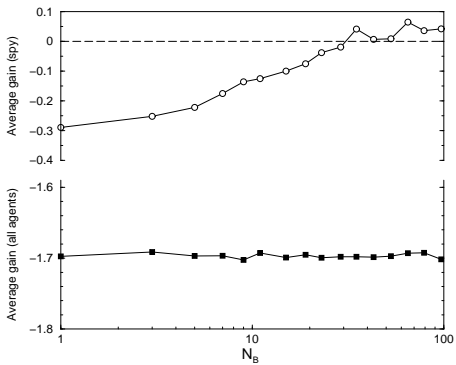,width=6cm}}
\caption{Gain of a spy versus the number of spied agents ($N=1001$, $\alpha=0.15$, average over 1000 realizations)}
\label{brokerNs}
\end{figure}

Since the outcome is anti-persistent in the symmetric phase and persistent in the asymmetric phase, only at the critical point there is no long term correlation in the outcome \cite{CM}. Accordingly, the spy always gain more than the average, except at the critical point where she gains the same (see figure \ref{brokeralpha}). With this setting, the agent has access in particular to the anti-persistence of the symmetric phase, explaining why even if only one agent is spied, the gain of the broker is much bigger (figure \ref{brokerNs}). 

Finally, the comparison between the two types of asymmetric information we have considered shows that it is much more interesting to spy than to have a larger memory : in the former case, one is sure to win more that the normal agents, except at the critical point.

\section{Conclusions}
In this work we have shown how to ask questions about real market mechanisms in a tay model. 
In spite of the severe simplification of the MG, with little modification one is able to study 
a broad spectrum of problems which could be dreamed of previously. The central result is to show 
that agents with limited rationality (or limited information processing power) can only make a 
market marginally efficient. To the first approximation one can say that these inductive players 
can maintain an approximate equilibrium, which is the central result of the El-Farol model. 
But studying carefully the fluctuations one finds that the fact that the market is more or 
less efficient does not imply that one can stop playing and sit at a randomly chosen site. 
Doing so would make the model less efficient. It is around this residual (marginal) 
inefficiency that the players are busy about.

With the introduction of producers the game can be of positive sum. We have shown how producers 
and speculators live in a symbiosis: producers are passive players who do not try to switch 
strategies. The reason is that they volontierly 
give up the speculation opportunities because 
they have outside business in mind. Thus producers inject information that the eager speculators 
are just happy to feed on. The speculators, while making away profits, perform a social function 
by providing liquidity thus reducing producers' market impact. We believe this is also true 
in real markets. Numerous other results show that it is now possible to systematically study 
markets with heterogeneous agents, whith real questions in mind.

\appendix

\section{Geometric and algebraic approaches to the MG}
\label{geom_alg}

This appendix is devoted to giving intuitive but rigorous views of what happens in the MG.

\subsection{Geometric approach to the MG}

The global behavior of the MG, measured by $\sigma^2$, can be quite well understood with a geometrical approach. Indeed, it is directly related to a much more intuitive geometrical concept : the Hamming distance between agents \cite{CZ2}, which is defined as follows for agents $i$ and $j$:
\be
\ovl{d_{i,j}}=\ovl{\frac{(a_{i}-a_{j})^2}{4}}=\frac{1}{2}-\frac{1}{2}\ovl{a_i a_j}.
\ee
It is worthwhile to note that $1-\ovl{d_{i,j}}$ equals the probability that both agents take the same action for a randomly drawn history, so for an agent, maximizing her distance with respect to all other agents is equivalent to maximizing her gain. Since the game is dynamical, one has to consider the time average of the actual Hamming distance between those two agents
\be
\ovl{\avg{d_{i,j}}}=\frac{1}{2}-\frac{1}{2}\ovl{\avg{a_i a_j}}.
\ee
The average Hamming distance per agent is then
\be\label{davg}
\ovl{\avg{d}}=\frac{1}{N(N-1)}\sum_{i,j}\ovl{\avg{d_{i,j}}}=\frac{1}{2}-\frac{1}{2N(N-1)}\sum_{i,j}\ovl{\avg{a_i a_j}}
\ee
The relationship between the distance and the fluctuations arises naturally by rewriting the latter as
\be\label{s2corr}
\sigma^2=N+\sum_{i\ne j}\ovl{\avg{a_i a_j}},
\ee
that is, as a sum over random fluctuations and correlations. Putting Eqs \req{davg} and \req{s2corr} together, one finds
\be\label{s2d}
\frac{\sigma^2}{N}=1-2(N-1)\pr{\frac{1}{2}-\ovl{\avg{d}}}.
\ee

This equation\footnote{It is exact for any $S$; even more, it remains exact if agents have not the same number of strategies.} links the geometrical \cite{CZ2,johnson} and the analytical approaches \cite{CMZe,CM}. It states that finding the average Hamming distance between the agents is equivalent to determining $\sigma^2$ by the analytical tools used in \cite{CMZe,MCZ,CM}. In general, it is impossible to find the average distance with a geometrical approach due to the fact that the Hamming distance is not transitive\footnote{The knowledge of $d_{i,j}$ and $d_{i,k}$ does not allow that of $d_{j,k}$.}. However, in the so-called reduced space of strategies (RSS) \cite{CZ2}, the distance is transitive, consequently Johnson et al. could find an approximate analytical expression of $\ovl{\avg{d}}$, and, by implicitly using Eq. \req{s2d} (which is straightforward in the RSS), they also gave an approximative expression of $\sigma^2$ \cite{johnson}. 
An equation quite similar to Eq. \req{s2d} also appears in \cite{MGNeural}, where it is shown that perceptrons 
playing the MG can cooperate.

\subsection{Algebraic approach to the phase transition}
\label{alg_appr}

We expose the algebraic origin of the phase transition. As it has been recalled, the 
agents actually try to minimize the available information $H$\cite{CMZe,MCZ}, and can actually cancel 
it when $\alpha<\alpha_c$. Let us see why. Since $H$ is a sum of $P$ non negative 
averages $\ovl{\avg{A}^2}$, $H=0$ only if all averages are zero, namely $\ovl{\avg{A}}=0~\forall \mu$, or equivalently 

\be
\sum_{i=1}^N\xi_i^\mu\avg{s_i}=-\Om^\mu~~~\forall \mu
\label{lin_sys}
\ee

These are $P$ linear equations in $N$ variables. However the $N$ variables $m_=\avg{s_i}$ are restricted to 
the $[-1,1]$ interval. Above $\alpha_c$ there are $N\phi$ variables which are frozen at the boundary of
this interval ($m_i=\pm 1$). Therefore there are $(1-\phi)N$ free variables only. 
As shown in refs. \cite{CMZe,MCZ}, the point $\alpha_c$ marks the transition below which the system of 
equations \req{lin_sys} becomes degenerate, i.e. when there are more variables than equations. Exactly
at $\alpha_c$ the number of free variables $(1-\phi)N$ exactly matches the number of equations $P$.
Dividing this equation by $N$ gives an equation for $\alpha_c$,
\be
\alpha_c=1-\phi
\ee
which is indeed confirmed nuerically to a high accuracy.

When $\alpha<\alpha_c$, there are much more free variables ($N$ indeed) than equations: 
the solutions of Eqs. \req{lin_sys} then belong to a subspace of dimension $N-P$. 
This allows the anti-persistent behavior to take place, because the system is free to move 
on this subspace. 
In the special case $c=0$, since there is no bias $\Omega^\mu =0$. 
The linear system of equation is then homogeneous and the solution $\avg{s_i}=0$ for all $i$ always exists. 
In particular, if $\alpha>1$, this solution is unique, hence $\sigma^2/N=1$. 
When $\alpha<1$, a subspace of solutions of dimension $N-P$ arises, and the 
anti-persistent behavior also takes place. Note that in this case, the system 
is always in the symmetric phase, therefore there is no phase transition. 

This argument easily generalizes to $S>2$ strategies\cite{MCZ}. If agents use, on average,
$n(S)$ strategies (and $S-n(S)$ are never used) the number of free variables is 
$N n(S)$. There are $P$ plus $N$ equations which these have to satisfy, where the
latter $N$ comes from the normalisation condition on the frequency with which each strategy is
used. At the critical point, these two numbers are equal, and we find 
\be\label{n_c-alpha_c}
n_c(S)=\alpha_c(S)+1.
\ee
At the critical point nearly one half of the strategies yield positive virtual gain
and are used, whereas the others are not used\cite{MCZ}. From this we find
\be
\alpha_c(S)=\alpha_c(2)+\frac{S}{2}-1
\ee

This shows that actually $\alpha_c$ grows linearly with $S$, but in a slightly less simple way that the 
one previously believed \cite{ZEnews,CZ2,CM,johnson}. 

Let us now show how the behavior of the agents is related to 
persistence/anti-persistence. We define $W$ as the average over the agents 
of $v_i^2$:
\bea
W&=&\frac{1}{N}\sum_{i=1}^N \pr{v_i}^2\\
&=&\lim_{T\to \infty}\frac{1}{T^2}
\sum_{t,t'=1}^{T}\sum_{i=1}^N\xi_i^{\mu(t)}\xi_i^{\mu(t')}A(t)A(t')
\eea
since the $\xi_i^\mu$ are independently drawn, 
\be
\frac{1}{N} \sum_{i=1}^{N}\xi_i^{\mu}\xi^{\mu}=
(1-c)\delta_{\mu,\mu}+O(1/\sqrt{N}).
\ee
Thus for large $N$ 

\be\label{apers}
W\simeq\lim_{T\to \infty}\frac{1}{T}\sum_{\tau=1}^T\overline{\avg{A(t)A(t-\tau)|\mu(t)=\mu(t-\tau)}}
\ee
where $\ovl{\avg{A(t)A(t-\tau)|\mu(t)=\mu(t')}}$ means that the average is taken over time 
for all $t$ and $\tau$ such that $\mu_t=\mu_{t-\tau}$, and summed over all histories. 
A closely realated quantity was first studied in ref. \cite{CM} where it was shown to
quantify anti-persistence in the symmetric phase.

Note that this equation implies that there can be no frozen agents unless the outcome exhibits 
persistence (i.e. $\overline{\avg{A(t)A(t-\tau)|\mu(t)=\mu(t-\tau)}}\neq 0$), 
which agrees with the analysis of \cite{CM}. In this case we find
\be\label{WH}
W\simeq \frac{H}{N}
\ee

On the other hand, no agent can freeze unless $H$ is non zero, 
which is the same as saying that the outcome is persistent.
Furthermore, the condition of freezing $v_i\ne0$ is equivalent to \cite{CM}

\be\label{frozencond}
\ovl{\xi_i\xi_i}<\left|\tilde h_i\right|,
\ee

where $\tilde h_i=\ovl{\Omega\xi_i}+\sum_{j\ne i}\ovl{\xi_i\xi_j}\avg{s_j}$.
It is worthwhile to see that $\ovl{\xi_i\xi_i}$ is the internal hamming distance. 
Eqs \req{apers} and \req{WH} give global conditions whether 
there can be frozen players or not, while Eq. \req{frozencond} 
give conditions on individual freezing. 

\section{The MG in biology} 

The MG model has another important application in biology: the sex ratio of 50:50. In the widely read
book of Richard Dawkins "Selfish Gene" \cite{Biology}, the Fischer theory  was brilliantly explained: if in the
offspring
pool either males or females were in minority, reproductive strategies for giving birth to a member in that
minority would enjoy a genetic advantage linearly proportional to the deviation from the 50:50 ratio.
The stable ratio is thus dynamically maintained. Brian Arthur's "El Farol" model, is also of the same 
genre, to show that using alternative strategies can lead to equilibrium. MG goes one step further:
while the equilibrium point is previously solved in different contexts by Fisher, Arthur et al, we
concentrate
on more refined questions.

\section{Replica method for the MG}
\label{replica}

For the sake of generality, we consider three different 
population of agents: 
\begin{enumerate}
\item the first population is composed of $N$ {\bf speculators}.
These are adaptive agents and they have each two {\em speculative}
strategies $a_{\up ,i}^\mu$, $a_{\down ,i}^\mu$ 
for $i=1,\ldots,N$ and $\mu=1,\ldots ,P$. These are 
drawn at random from the pool of all strategies, independently
for each agent. We allow a correlation among the two
strategies of the same agent:
\bea
P(a_\up,a_\down)&=&\frac{c}{2}\left[\delta_{a_\up,+1}
\delta_{a_\down,+1}+
\delta_{a_\up,-1}\delta_{a_\down,-1}\right]\nonumber\\
&+&\frac{1-c}{2}\left[\delta_{a_\up,-1}\delta_{a_\down,+1}+
\delta_{a_\up,-1}\delta_{a_\down,+1}\right]
\eea
Note that, for $c=0$ agents choose just one strategy $a_\up$ 
and fix $a_\down=-a_\up$ as its opposite, whereas for $c=1$ they
have one and the same strategy $a_\up=a_\down$. The original 
random case \cite{CZ1,Savit} corresponds to $c=1/2$.
These agents assign scores $U_{s,i}(t)$ to each of their 
strategies and play the strategy $s_i(t)$ with the highest score,
as discussed in the text. Therefore for speculators:
\be
a_{\rm spec}(t)=a^{\mu(t)}_{s_i(t),i}.
\ee
\item then we consider $N_{\rm prod}^{\rm indep}=\rho N$ 
{\bf producers}:
They have only one randomly and independently drawn strategy 
$b_i^\mu$ so 
\be
a_{\rm prod}(t)=b^{\mu(t)}_i.
\ee
Producers have a predictable behavior in the market and they are
not adaptive. Instead of $\rho N$ {\em independent} producers
one can also consider $N_{\rm prod}^{\rm dep}$ correlated producers
who all have the same predictable behavior $b_{\rm prod}^\mu$.
\item Finally we consider $\eta N$ {\bf noise traders}. These 
are defined as agents whose actions are given by
\be
a_{\rm noise}(t)={\rm random sign}.
\ee
Each noise trader as a random number generator which is
independent of each other agent.
\end{enumerate}

It has been shown \cite{CMZe,MCZ} that the stationary state 
properties of the MG
are described by the ground state of $H$. Note that this approach fails 
however to reproduce the anti-persistent behavior which is at the
origin of crowd effects in the symmetric phase. 
In our case 
\be
A(t)=A_{\rm spec}(t)+A_{\rm prod}(t)+A_{\rm noise}(t)
\ee
where
\be
A_{\rm spec}(t)=\sum_{j=1}^Na^{\mu(t)}_{s_j(t),j}
\ee
and
\be
A_{\rm prod}(t)=\sum_{j=1}^{\rho N} b^{\mu(t)}_j\equiv 
A_{\rm prod}^{\mu(t)}
\label{Aprod}
\ee
and $A_{\rm noise}(t)=2k(t)-\eta N$ where $k(t)$ is a binomial
random variable with $P(k)={\eta N \choose k} 2^{-\eta N}$.
Since $H=\ovl{\avg{A}^2}$ and the contribution of noise traders
to $\avg{A^\mu}$ vanishes $\avg{A_{\rm noise}}=0$, the collective
behavior of the system is independent of $\eta$. Noise traders 
shall contribute a constant $\eta N$ to $\sigma^2$ and will not
affect other agents. This only holds in the asymmetric phase
(see text). We can then reduce to the study of speculators and
producers only. 

Let us define,
for convenience, $A^\mu=A^\mu_{\rm spec}+\lambda A^\mu_{\rm prod}$
where 
\be
A^\mu_{\rm spec}=
\sum_{i=1}^N \left[a_{\up,i}^\mu\frac{1+s_i}{2}+
a_{\down,i}^\mu \frac{1-s_i}{2}\right]
\ee
and $A^\mu_{\rm prod}$ is given in Eq. \req{Aprod}. 
Here $s_i$ is the dynamical variable controlled by speculator $i$.
We shall implicitly consider directly time averaged 
quantities so $s_i$ is a real variable in $[-1,1]$ rather than
a discrete one.
The parameter $\lambda$ is inserted so that, once we have computed
the energy $H=\overline{(A_{\rm spec}+\lambda A_{\rm prod})^2}$ we 
can compute the total gain $G_{\rm prod}$ of producers by
\[
G_{\rm prod}\equiv -\overline{A A_{\rm prod}}=-
\left.\frac{1}{2}\frac{\partial H}{\partial \lambda}\right|_{\lambda=1}.
\]
The gain of speculators is obtained subtracting this contribution and 
that of noise traders from the total gain $-\sigma^2$
\be
G_{\rm spec}=-\sigma^2+\eta N-G_{\rm prod}.
\label{gainspecap}
\ee

\subsection{Replica calculation}

The zero temperature behavior of the Hamiltonian $H$ can be studied
with spin glass techniques \cite{dotsenko,MPV}. We introduce $n$ replicas
of the system, each with dynamical variables $s_{i,c}$, 
labeled by replica indices $c,d=1,\ldots,n$. Then we write
replicated partition function:
\be
\avg{Z^n(\beta)}=\Tr_s\prod_{\mu,c}\Avg{e^{-\frac{\beta}{P}
(A_c^\mu)^2}}_{a,b}
\label{Znb}
\ee
where the average is over the disorder variables
$a_{s,i}^\mu$, $b_i^\mu$ and $\Tr_s$ is the trace on the variables 
$s_{i,c}$ for all $i$ and $c$.
Following standard procedures \cite{dotsenko,MPV}, we 
introduce a Gaussian variable $z_c^\mu$ so that we can linearize 
the exponent in Eq. \req{Znb}. This allows us to carry out the 
averages over $a$'s and $b$'s explicitly. Then we introduce 
new variables $Q_{c,d}$ and $r_{c,d}$ with the identity
\beas
1&=&\int dQ_{c,d} \delta \left( Q_{c,d}-\frac{1}{N}
\sum_i s_{i,d}s_{i,d}\right)\\
&\propto &
\int dr_{c,d}dQ_{c,d}e^{-\frac{\alpha\beta^2}{2}
r_{c,d}\left( N Q_{c,d}-\sum_i s_{i,c}s_{i,d}\right)}
\eeas
for all $c\ge d$, which allow us to write the partition function
(to leading order in $N$) as:
\beas
\Avg{Z^n(\beta)}= \int d\hat{Q}d\hat{r}
e^{-Nn\beta F(\hat Q,\hat r)}
\eeas
with
\bea
F(\hat Q,\hat r)&=&\frac{\alpha}{2n\beta}\Tr\log \hat T+
\frac{\alpha\beta}{2n}\sum_{c\le d}r_{c,d}Q_{c,d}\nonumber\\
&-&\frac{1}{n\beta}\log\left[\Tr_s e^{\frac{\alpha\beta^2}{2}
\sum_{c\le d}r_{c,d}s_c s_d}\right].
\eea
The matrix $\hat T$ is given by
\[
T_{a,b}=\delta_{a,b}+\frac{2\beta}{\alpha}\left[c+\rho+(1-c)
Q_{a,b}\right].
\]
For {\em correlated} producers we would have obtained the same
result but with $\rho\to\rho+\rho^2 N \epsilon^2$, where 
$\epsilon$ measures the bias of producers towards a particular 
action for a given $\mu$, or equivalently the correlation between 
the actions of two distinct producers. 
More precisely $\epsilon^2$ is the average of $b_i^\mu b_j^\mu$ 
for $i\neq j$ and for all $\mu$.
Therefore the limit $\rho\to\infty$ also corresponds to a small 
share of producers $\rho\ll 1$ with a small bias $\epsilon\not = 
0$. Note that a bias $\epsilon\sim\sqrt{N}$ corresponds 
indeed to $\sim N$ independent producers. Equivalently 
$\sim\sqrt{N}$ correlated producers, with $\epsilon$ finite
are equivalent to $\sim N$ independent producers.

With the replica symmetric ansatz
\[
Q_{c,d}=q+(Q-q)\delta_{c,d},~~~~~~r_{c,d}=2r+(R-2r)\delta_{c,d}
\]
the matrix $\hat T$ has $n-1$ degenerated eigenvalues 
$\lambda_0=1+\frac{2(1-c)\beta(1-q)}{\alpha}$ and one eigenvalue
equal to $\lambda_1=\frac{2\beta[c+\rho+(1-c)q]}{\alpha}n+1+
\frac{2(1-c)\beta(1-q)}{\alpha}$
therefore, after standard algebra,
\bea
F^{(RS)}(q,r)&=&\frac{\alpha}{2\beta}
\log\left[1+\frac{2(1-c)\beta(Q-q)}{\alpha}\right]\nonumber\\&+&
\frac{\alpha[c+\rho+(1-c)q]}{\alpha+2(1-c)\beta(Q-q)}+
\frac{\alpha\beta}{2}(RQ-rq)\nonumber\\
&-&\frac{1}{\beta}
\Avg{\log\int_{-1}^1 ds e^{-\beta V_z(s)}}
\label{FRS}
\eea
where we found it convenient to define the ``potential''
\be
V_z(s)=-\frac{\alpha\beta(R-r)}{2} s^2-\sqrt{\alpha r}\,z\,s
\ee
so that the last term of $F^{(RS)}$ looks like the free
energy of a particle in the interval $[-1,1]$ with potential
$V_z(s)$ where $z$ plays the role of disorder. 

The saddle point equations are given by:
\bea
\frac{\partial F^{(RS)}}{\partial q}=0 ~~~ &\Rightarrow & ~~~~
r=\frac{4(1-c)[c+\rho+(1-c)q]}{[\alpha+2(1-c)\beta(Q-q)]^2}\\
\frac{\partial F^{(RS)}}{\partial Q}=0 ~~~  &\Rightarrow & ~~~~
\beta(R-r)=-\frac{2(1-c)}{\alpha+2 (1-c) \beta(Q-q)}\\
\frac{\partial F^{(RS)}}{\partial R}=0 ~~~  &\Rightarrow & ~~~~
Q=\avg{\avg{s^2}}\\
\frac{\partial F^{(RS)}}{\partial r}=0 ~~~  &\Rightarrow & ~~~~
\beta(Q-q)=\frac{\avg{\avg{sz}}}{\sqrt{\alpha r}}
\eea
where $\avg{\avg{\cdot}}$ stands for a thermal average over the
above mentioned one particle system.

In the limit $\beta\to 0$ we can look for a solution with
$q\to Q$ and $r\to R$. It is convenient to define 
\be
\chi=\frac{2(1-c)\beta(Q-q)}{\alpha},~~~\hbox{and}~~~
\zeta=-\sqrt{\frac{\alpha}{r}}\beta(R-r)
\ee
and to require that they stay finite in the limit $\beta\to\infty$.
The averages are easily evaluated since, in this case, they are 
dominated by the minimum of the potential 
$V_z(s)=\sqrt{\alpha r}(\zeta s^2/2-zs)$ for 
$s\in [-1,1]$. The minimum
is at $s=-1$ for $z\le -\zeta$ and at $s=+1$ for $z\ge \zeta$.
For $-\zeta<z<\zeta$ the minimum is at $s=z/\zeta$. With this
we find
\be
\avg{\avg{sz}}=\frac{1}{\zeta}
\erf\left(\frac{\zeta}{\sqrt{2}}\right)
\ee
and
\be
\avg{\avg{s^2}}=Q=1-\sqrt{\frac{2}{\pi}}\frac{e^{-\zeta^2/2}}{\zeta}
-\left(1-\frac{1}{\zeta^2}\right)
\erf\left(\frac{\zeta}{\sqrt{2}}\right)
\label{eqQ}
\ee
With some more algebra, one easily finds:
\be
\chi=\left[\alpha/\erf\left(\frac{\zeta}{\sqrt{2}}\right)-1\right]^{-1}
\label{eqx}
\ee
Finally $\zeta$ is fixed as a function of $\alpha$ by the equation
\be
\sqrt{\frac{2}{\pi}}\frac{e^{-\zeta^2/2}}{\zeta}+
\left(1-\frac{1}{\zeta^2}\right)
\erf\left(\frac{\zeta}{\sqrt{2}}\right)
+\frac{\alpha}{\zeta^2}=\frac{1+\rho}{1-c}
\label{eqzeta}
\ee
Note that $\zeta$ only depends on the combination $(1+\rho)/(1-c)$
which runs from $1$ -- for $\rho=c=0$ i.e. no producers and ``perfect'' 
speculators -- to $\infty$. The latter limit
occurs either if $c\to 1$, i.e. when speculators become producers, or 
if $\rho\to\infty$ (many producers).

Eq. \req{eqx} means that $\chi$ diverges when $\alpha\to\alpha_c(\rho,c)^+$,
which then implies that at the critical point
\be
\erf\left(\frac{\zeta}{\sqrt{2}}\right)=\alpha=\alpha_c.
\ee
This back in the other saddle point equations, yields
the following equation for $\zeta=\zeta_c$:
\be
\sqrt{\frac{2}{\pi}}\frac{e^{-\zeta^2/2}}{\zeta}
+\erf\left(\frac{\zeta}{\sqrt{2}}\right)
=\frac{1+\rho}{1-c}.
\ee

The free energy, at the saddle point, for $\beta\to\infty$, is
\be
F^{(RS)}=
\frac{c+(1-c)Q+\rho}{(1+\chi)^2}
\ee
where $Q$ and $\chi$ take their saddle point values Eqs. \req{eqQ} and
\req{eqx}. 

The gain of producers, from Eq. \req{FRS}, is
\be
\frac{G_{\rm prod}}{N}=-\frac{\rho}{1+\chi}
\ee
and that of speculators is obtained from Eq. \req{gainspecap}.

At $\alpha_c$ $\chi\to\infty$ so that $F^{(RS)}\to 0$.
Note that the loss of producers vanishes $L_{\rm prod}\to 0$ as
$\alpha\to\alpha_c$, whereas the loss of speculators
$L_{\rm spec}=(1-Q)/2$ is always positive below $\alpha_c$. 

The phase diagram is shown in figure \ref{phasediagr}. here we
discuss some limits.

\section{The sign MG}
\label{sign}
The original MG \cite{CZ1} is defined with payoffs
\be
g_i=-a_i(t)\sign A(t)
\ee
Over a long period of time $T$, the change in $\Delta_i(t)$
is given by
\bea
&&\frac{\Delta_i(t+T)-\Delta_i(t)}{T}=\frac{1}{T}
\sum_{\tau=t}^{T-1}\xi_i^{\mu(\tau)}\sign A(\tau)\nonumber\\
&&\ \ \simeq \frac{1}{P}\sum_{\mu=1}^P 
\left[2{\rm Prob}\{A(\tau)>0|\mu(\tau)=\mu\}-1\right]
\xi_i^{\mu}
\eea

Then, for any fixed $\mu$, the relevant quantity is the
probability that $A(t)>0$, when $\mu(t)=\mu$. This can be 
computed within our mean-field approximation: Indeed
if $\avg{s_i}=m_i$ we can regard $s_i$ as a random variable
with distribution 
\[
P(s_i=\pm 1)=\frac{1\pm m_i}{2}.
\]
Then, the relation 
$A^{\mu(t)}(t)=\Om^\mu+\sum_{i=1}^N\xi_i^\mu s_i$
implies that we can consider $A^\mu$ as a Gaussian 
variable with variance 
\[
{\rm Var}(A^\mu)=\sum_{i=1}^N {\xi_i^\mu}^2 (1-m_i^2)+\eta N
\]
This allows us to compute 
\[
{\rm Prob}\{A(\tau)>0|\mu(\tau)=\mu\}=\frac{1}{2}{\rm erfc}
\left(\frac{\avg{A^\mu}}{\sqrt{2{\rm Var}(A^\mu)}}\right).
\]
Note that close to the critical point $\alpha_c$, $\avg{A^\mu}$ is
very small compared to ${\rm Var}(A^\mu)$, which means that it
is legitimate to expand the ${\rm erfc}$ function to linear
order. This gives us back a linear minority game, but with 
\be
g_i=-\frac{a_i(t) A(t)}{\sqrt{2{\rm Var}(A^\mu)}}.
\ee
Note, then that when $\eta$ increases the gains for each
speculator decreases. This is actually true even away from 
$\alpha_c$. It is indeed easy to check that 
$\avg{\sign A(t)}$ decreases as $\eta$ increases.





\end{document}